\documentclass{article}
\usepackage{graphicx} 
\usepackage{float}
\usepackage{tikz}
\usepackage{bm}
\usepackage{gensymb}
\tikzstyle{startstop} = [rectangle, rounded corners, minimum width=3cm, minimum height=1cm,text centered, draw=black, fill=red!30]

\tikzstyle{io} = [trapezium, trapezium left angle=70, trapezium right angle=110, minimum width=3cm, minimum height=1cm, text centered, draw=black, fill=blue!30]

\tikzstyle{process} = [rectangle, minimum width=3cm, minimum height=1cm, text centered, draw=black, fill=orange!30]

\tikzstyle{decision} = [diamond, minimum width=3cm, minimum height=1cm, text centered, draw=black, fill=green!30]

\tikzstyle{arrow} = [thick,->,>=stealth]
\tikzstyle{process} = [rectangle, minimum width=3cm, minimum height=1cm, text centered, text width=3cm, draw=black, fill=orange!30]
\usetikzlibrary{shapes.geometric, arrows}
\usepackage{comment}
\usepackage[ruled,vlined,noline,boxed, noend]{algorithm2e}
\usepackage{subcaption}
\usepackage{hyperref}
\usepackage{setspace}
\usepackage{epsfig}  
\usepackage{graphicx}  
\usepackage{amsmath}  
\usepackage{amssymb}  
\usepackage{lscape}  
\usepackage[justification=raggedright]{caption}	

\usetikzlibrary{decorations.pathreplacing}
\usetikzlibrary{calc,patterns,decorations.pathmorphing,decorations.markings}
\title{High-Performance Gradient Evaluation for Complex Soft Materials Using MPI-based DFS Algorithm}
\author{Anurag Bhattacharyya}
\author{Anurag Bhattacharyya\footnote{Lawrence Livermore National Laboratory, bhattacharyy3@llnl.gov}}
\date{}

\begin{document}

\maketitle
\begin{abstract}

\noindent This article presents a depth-first search (DFS)-based algorithm for evaluating sensitivity gradients in the topology optimization of soft materials exhibiting complex deformation behavior. The algorithm is formulated using a time-dependent adjoint sensitivity approach and is implemented within a PETSc-based C++ MPI framework for efficient parallel computing. It has been found that on a single processor, the sensitivity analysis for these complex materials can take approximately 45 minutes. This necessitates the use of high-performance computing (HPC) to achieve feasible optimization times. This work provides insights into the algorithmic framework and its application to large-scale generative design for physics integrated simulation of soft materials under complex loading conditions.

\end{abstract}
\section{Introduction}
Shape memory polymers (SMPs) are a class of active polymeric materials that have the ability to regain their original undeformed configuration from a deformed state under the application of an external stimulus. A time-dependent adjoint sensitivity formulation implemented through a recursive algorithm is used to calculate the gradients required for the topology optimization algorithm.
For 3D SMPTO code, ran on UIUC campus-cluster, the time taken for single optimization iteration amounts to approximately 10 hours on 200 processors for a 3D mesh of 50$\times$ 20 $\times$ 20 elements with 60, 000 degrees-of-freedom. Our aim is to get the maximum number of optimization iterations completed within a specific unit of time, which would demand resources beyond the current allowable limit of the UIUC campus-cluster. The code is provided in the  github repository\cite{githubGitHubBhttchr6TO_ShapeMemoryPolymer}. This article provides a detailed explanation of the PETSc-based algorithm implemented in thesis\cite{bhattacharyya2021hierarchical} to provide interested readers with additional background information.
\begin{figure}[H]
	\centering
	\includegraphics[width=8cm,trim={0cm 0cm 0cm 0cm },clip]{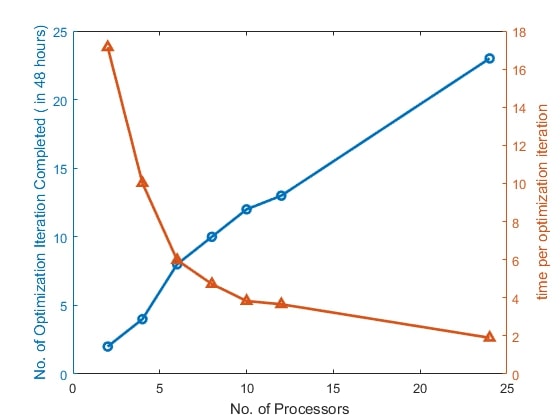}
	\caption{Scaling capability study performed using UIUC campus-cluster}
	\label{sens domain}
\end{figure}
\section{Sensitivity Analysis}
\label{Time-dependent Adjoint Sensitivity Analysis}
Time-dependent adjoint sensitivity analysis is performed to calculate the gradient information required for the structural optimization process. The procedure here describes the calculation of adjoint sensitivities. The function of interest being differentiated is the displacement at a particular degree-of-freedom ($a$) of the structure, at a particular time step ($M$) as shown in Figure (\ref{sens domain}).
\begin{figure}[H]
	\centering
	\includegraphics[width=8cm,trim={0cm 0cm 0cm 0cm },clip]{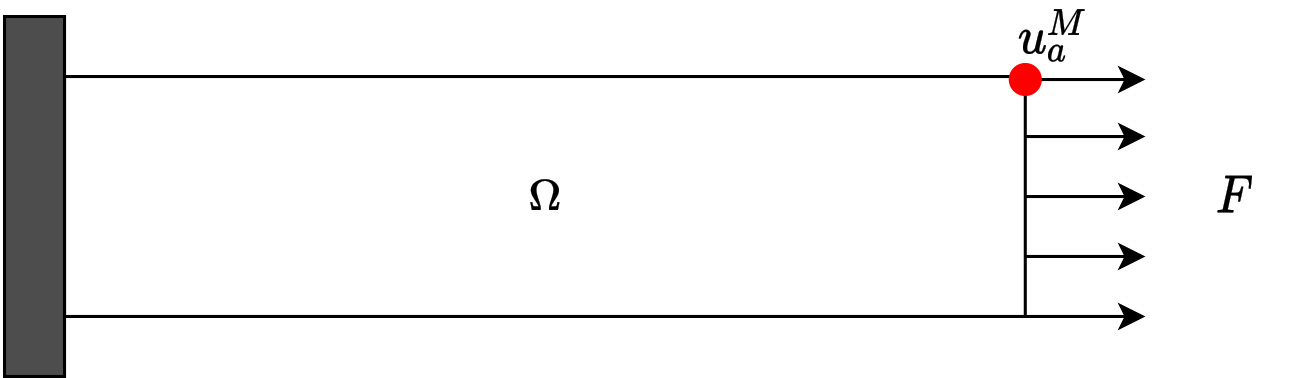}
	\caption{Design domain for sensitivity calculations and its verification.}
	\label{sens domain}
\end{figure}
\noindent Let the scalar function of interest($\theta$) be defined as 
\begin{align}
	\theta={\bm{u}_a^M(\rho)}
	\label{objective function}
\end{align}
Let $\bm{u^M(\rho)}$ represent the displacement vector of the whole structure at time step $M$. Then we can write Equation(\ref{objective function}) as
\begin{align}
	\theta=\bm{L}^T\bm{u}^M(\rho)
	\label{fil obj}
\end{align}	
where $\bm{L}$ is a column vector and is zero everywhere except at the entry corresponding to the  $a^{th}$ degree-of-freedom. We can form an augmented Lagrangian function as
\begin{align}
	\Theta=\theta+\sum_{i=1}^{M}\Bigg[ {\bm{\lambda}^{(i)}}^T \bm{R}^{(i)}(\bm{\rho}, \bm{u}^i, \bm{u}^{i-1},....,\bm{u}^0)\Bigg]
	\label{lagrangian}
\end{align}
where $\bm{\rho}$ is the design variable and the variable $\bm{u}$ is the state variable(containing all the variables evaluated through forward analysis). Note that $\Theta =\theta$ since $R^{(i)}=0$ for all i. Therefore $\frac{d\Theta}{d\mathbf{\rho}}=\frac{d\theta}{d\mathbf{\rho}}$. Differentiating Equation (\ref{lagrangian}) with respect to the design variable $\bm{\rho}$, we obtain
\begin{align}
	\frac{d\Theta}{d\bm{\rho}}&=L^T\frac{d\bm{u}^M}{d\bm{\rho}}+\sum_{i=1}^{M}\Bigg[{\bm{\lambda}^{(i)}}^{T}\Bigg(\sum_{k=1}^{i}\frac{\partial \bm{R}^{(i)}}{\partial \bm{\bm{u}}^{(k)}}\frac{d\bm{u}^{(k)}}{d\bm{\rho}}+\frac{\partial \bm{R}^{(i)}}{\partial \bm{\bm{\rho}}}\Bigg)\Bigg]
	\label{eq 18}
\end{align}
Expanding the right-hand side terms yields
\begin{align}
	\begin{split}
		\frac{d\Theta}{d\bm{\rho}}&=L^T\frac{d\bm{u}^M}{d\bm{\rho}}+\sum_{i=1}^{M}{\bm{\lambda}^{(i)}}^T\frac{\partial \bm{R}^{(i)}}{\partial \bm{\bm{\rho}}}+{\bm{\lambda}^{(M)}}^T\Bigg(\frac{\partial \bm{R}^{(M)}}{\partial \bm{u}^{(M)}}\frac{d\bm{u}^{(M)}}{d\bm{\rho}}\Bigg)+\sum_{i=1}^{M-1}\sum_{k=i}^{M}\Bigg({\bm{\lambda}^{(k)}}^T\frac{\partial \bm{R}^{(k)}}{\partial \bm{\bm{u}}^{(i)}}\Bigg)\frac{d\bm{u}^{(i)}}{d\bm{\rho}}\\
	\end{split}
\end{align}
The solution of \{$\bm{\lambda}^i$\} which causes all the implicit terms, $\{\frac{d\bm{u}}{d\bm{\rho}}\}$\footnote{Note that implicit derivatives, $\frac{d*}{d\bm{\rho}}$, capture implicit dependence of a function or state variable with respect to $\rm{\rho}$ due to the solution of the residual, whereas explicit derivatives capture only direct dependence.  Consequently, implicit derivatives are more expensive to evaluate, and therefore we seek to eliminate them from the sensitivity calculation}, to vanish is given by 
\begin{align}
	\begin{split}
		\bm{\lambda}^{(M)}&=-L^T\Bigg[\frac{\partial \bm{R}^{(M)}}{\partial \bm{u}^{(M)}}\Bigg]^{-1}\\
		\bm{\lambda}^{(i)}&=-\Bigg[\sum_{k=i+1}^{M}{\bm{\lambda}^{(k)}}^T\frac{\partial \bm{R}^{(k)}}{\partial \bm{u}^{(i)}}\Bigg]\Bigg[\frac{\partial \bm{R}^{(i)}}{\partial \bm{u}^{(i)}}\Bigg]^{-1}\\\
	\end{split}
	\label{adjoint vectors}
\end{align}
When solved in this way the parameters \{$\bm{\lambda}^i$\} are referred to as the $\it{adjoint}$ vectors, and each vector $\bm{\lambda^i}$ represents the adjoint state at each time step $t_i$.  \href{Algo_1}{Algorithm 1} contains a pseudocode description of the algorithm used to compute the sensitivities of the SMP material.

\begin{algorithm}[H]
    \DontPrintSemicolon
    $\frac{df}{d\bm{\rho}} \leftarrow \frac{\partial}{d\bm{\rho}}\left[f(\bm{\rho}, {\bm{u}}^{(M)})\right]$ \tcp{initialize sensitivities}
    
    $\bm{\lambda}^{(M)}\leftarrow {\left[ \frac{\partial \bm{R}^{(M)}}{\bm{u}^{(M)}}\right]}^{-1} \left[ -\frac{\partial f}{\bm{u}^{(M)}}\right]$ \tcp{solve for final adjoint state}
    
    \For{$i\leftarrow M, M-1, \dots, 0$}{
        \tcp{cycle back through each time step}
        ${\bm{F}_{RHS}} \leftarrow 0$\; 
        \tcp{Cycle forward through all subsequent time steps}
        \For{$k \leftarrow i+1, i+2, \dots, M$}{
            $\frac{\partial{\bm{R}}^{(k)}}{\partial{\bm{u}}^{(i)}} \leftarrow \frac{\partial {\bm{\varepsilon}}^{ir}_{(k)}}{\partial \bm{\varepsilon}^r_{(i)}} + \frac{\partial {\bm{\varepsilon}}^{ig}_{(k)}}{\partial \bm{\varepsilon}^r_{(i)}} +\frac{\partial {\bm{\varepsilon}}^{i}_{(k)}}{\partial \bm{\varepsilon}^r_{(i)}} +\frac{\partial {\bm{\varepsilon}}^{is}_{(k)}}{\partial \bm{\varepsilon}^r_{(i)}}$\;
            \tcp{Each additive term is traced back in time through the recursive \href{algo_2}{Algorithm 2} and \href{algo_3}{Algorithm 3}}
            ${\bm{F}_{RHS}} \leftarrow {\bm{F}_{RHS}}-\left[ {\bm{\lambda}}^{(k)}\frac{\partial {\bm{R}}^{(k)}}{\partial {\bm{u}}^{(i)}}\right]$\;
        }
        ${\bm{\lambda}}^{(i)} \leftarrow -\bm{F}_{RHS}{\left[ \frac{\partial {\bm{R}}^{(i)}}{\partial{\bm{u}}^{(i)}}\right]}^{-1}$ \tcp{solve for intermediate adjoint vectors}
        $\frac{df}{d\bm{\rho}} \leftarrow \frac{\partial f}{\partial \bm{\rho}} + {\bm{\lambda}}^{(i)}\frac{\partial {\bm{R}}^{(i)}}{\partial \bm{\rho}}$\;
    }
    \caption{Time-dependent adjoint sensitivity analysis }
\end{algorithm}

Once we obtain the full set of \emph{adjoint} vectors, the sensitivities can be obtained as
\begin{align}
	\begin{split}
		\frac{d\Theta}{d\bm{\rho}}&=\sum_{i=0}^{M}\bm{\lambda}^i\frac{\partial \bm{R}^{i}}{\partial\bm{\rho}}	
		\label{sens eqn}
	\end{split}
\end{align}
\section{Derivation of sensitivity analysis}
Having discussed the generalized formulation for time-dependent adjoint sensitivity analysis in section \ref{Time-dependent Adjoint Sensitivity Analysis}, we focus on deriving the sensitivity formulation specifically for shape memory polymers. To avoid confusion in the notation representing inelastic strain components and time steps, from here on the current time step will be denoted by subscript \{$n+1$\}, the previous time step will be denoted by subscript \{$n$\} and so on.
\\ The sensitivity of the objective function is calculated via Equation {(\ref{sens eqn})}. This equation has two components, the first is the \emph{adjoint} vectors ($\bm{\lambda}$) and the other is the component capturing the explicit dependence of the residual term on the design variable. The \emph{adjoint} vectors are computed via Equation {(\ref{adjoint vectors})}. Evaluation of both of these components requires the residual term ($\bm{R}$). The residual equation for the SMP can be stated as
\begin{align}
	\begin{split}
		\bm{R}_{n+1}&=\int_{\Omega}^{} \bm{B}^T\mathbb{A}^{(r)}\mathbb{D}^{-1}_{n+1}:\bm{B}\bm{u}_{n+1} dv-\int_{\Omega}^{} \bm{B}^T\mathbb{X}^{(r)}_{n+1}:\bm{\varepsilon}^{(ir)}_{n} dv
		\\&+\int_{\Omega}^{} \bm{B}^T\mathbb{X}^{(g)}_{n+1}:\bm{\varepsilon}^{(ig)}_{n} dv
		+\int_{\Omega}^{} \bm{B}^T\mathbb{Y}^{(r)}_{n+1}:\bm{\varepsilon}^{(ir)}_{n} dv
		-\int_{\Omega}^{} \bm{B}^T\mathbb{V}_{n+1}:\bm{\varepsilon}^{(i)}_{n} dv
		\\&-\int_{\Omega}^{} \bm{B}^T\mathbb{Z}^{(r)}_{n+1}:\bm{\varepsilon}^{(is)}_{n} dv
		-\int_{\Omega}^{} \bm{B}^T\mathbb{A}^{(r)}\mathbb{D}^{-1}_{n+1}:\bm{\varepsilon}^{Th}_{n+1} dv-\bm{F}^{ext}\\
		\label{residual eqn}
	\end{split}
\end{align}
where the terms $\mathbb{X}^{(r)}_{n+1} $, $\mathbb{X}^{(g)}_{n+1} $, $\mathbb{Y}^{(r)}_{n+1} $, $\mathbb{V}^{(r)}_{n+1} $, $\mathbb{Z}^{(r)}_{n+1} $ are given by
\begin{align}
	\begin{split}
		\mathbb{X}^{(r)}_{n+1} &= \mathbb{A}_r\mathbb{D}^{-1}_{n+1}\phi^{(g)}_{n+1}{\mathbb{A}_g}^{-1}\mathbb{B}_r\\
		\mathbb{X}^{(g)}_{n+1} &= \mathbb{A}_r\mathbb{D}^{-1}_{n+1}\phi^{(g)}_{n+1}{\mathbb{A}_g}^{-1}\mathbb{B}_g\\
		\mathbb{Y}^{(r)}_{n+1} &= \mathbb{A}_r\mathbb{D}^{-1}_{n+1}\Bigg(\frac{\Delta t}{\eta_i}\Bigg)\mathbb{B}_r\\
		\mathbb{V}^{(r)}_{n+1} &= \mathbb{A}_r\mathbb{D}^{-1}_{n+1}\\
		\mathbb{Z}^{(r)}_{n+1} &= \mathbb{A}_r\mathbb{D}^{-1}_{n+1}\\
	\end{split}
\end{align}
The differentiation of the residual equation, $\bm{R}_{n+1}$,  with respect to the design variables can be computed by   
\begin{align}
	\begin{split}
		\frac{\partial\bm{R}_{n+1}}{\partial\bm{\rho}}&=\int_{\Omega}^{}\bm{B}\frac{\partial\bm{\sigma}_{n+1}}{\partial\bm{\rho}}dv-\frac{\partial\bm{F}^{ext}_{n+1}}{\partial\bm{\rho}}\\
		\frac{\partial \sigma_{n+1}}{\partial \bm{\rho}}&=\frac{\partial \mathbb{A}_{r}}{\partial \rho}:\bm{\varepsilon}^{(r)}_{n+1}+\mathbb{A}_r:\frac{\partial \bm{\varepsilon}^{(r)}_{n+1} }{\partial \bm{\rho}}-\frac{\partial \mathbb{B}_r}{\partial \bm{\rho}}:\bm{\varepsilon}^{(ir)}_{n}-\mathbb{B}_r:\frac{\partial \bm{\varepsilon}^{(ir)}_{n}}{\partial \bm{\rho}}\\
	\end{split}
	\label{drdx}
\end{align}
Let us evaluate these terms one by one. The tensor $\mathbb{A}_r$ is written as:
\begin{equation}
	\mathbb{A}_r = \mathbb{K}^r_{neq}+\mathbb{K}^r_{eq}-\frac{\Delta t}{\eta_r}\mathbb{K}^r_{neq}{\mathbb{H}^r}^{-1}\mathbb{K}^r_{neq}
\end{equation}
Derivative of this term with respect to the design variable can be written as :
\begin{multline}
	\frac{\partial \mathbb{A}_r}{\partial \bm{\rho}} = \frac{\partial \mathbb{K}^r_{neq}}{\partial \bm{\rho}}+\frac{\partial \mathbb{K}^r_{eq}}{\partial \bm{\rho}}-
	\frac{\Delta t}{\eta_r}\frac{\partial \mathbb{K}^r_{neq}}{\partial \bm{\rho}}{\mathbb{H}_r}^{-1}\mathbb{K}^r_{neq}
	+\frac{\Delta t}{\eta_r}\mathbb{K}^r_{neq}{\mathbb{H}_r}^{-1}\frac{\partial \mathbb{H}_r}{\partial \bm{\rho}}{\mathbb{H}_r}^{-1}\mathbb{K}^r_{neq}
	\\-\frac{\Delta t}{\eta_r}\mathbb{K}^r_{neq}{\mathbb{H}_r}^{-1}\frac{\partial\mathbb{K}^r_{neq}}{\partial \bm{\rho}}
	+\frac{\Delta t}{\eta^2_r}\frac{\partial \eta_r}{\partial \bm{\rho}}\mathbb{K}^r_{neq}{\mathbb{H}_r}^{-1}\mathbb{K}^r_{neq}
\end{multline}
The term for the glassy-phase $	\frac{\partial \mathbb{A}_g}{\partial \bm{\rho}}$ can be evaluated similarly. The derivative of the terms in the above equation is given below.
\begin{equation}
	\frac{\partial \mathbb{H}_r}{\partial \bm{\rho}} = \frac{\Delta t}{\eta_r}\frac{\mathbb{K}^r_{neq}}{\partial \bm{\rho}}-\frac{\Delta t}{\eta^2_r}\mathbb{K}^r_{neq}\frac{\partial \eta_r}{\partial \bm{\rho}}
\end{equation} 

The derivative of the rubbery phase strain with respect to the design variable can be written as:
\begin{multline}
	\frac{\partial \bm{\varepsilon}^r_{n+1}}{\partial \bm{\rho}} =\Bigg[-{\mathbb{D}_{n+1}}^{-1}\frac{\partial \mathbb{D}_{n+1}}{\partial \bm{\rho}}{\mathbb{D}_{n+1}}^{-1}\Bigg]:\bm{C}_{n+1} + {\mathbb{D}_{n+1}}^{-1}:\frac{\partial \bm{C}_{n+1}}{\partial \bm{\rho}}
\end{multline}
The term $\frac{\partial \mathbb{D}_{n+1}}{\partial \bm{\rho}}$ and $\frac{\partial \bm{C}_{n+1}}{\partial \bm{\rho}}$ can be evaluated as:
\begin{multline}
\frac{\partial \mathbb{D}_{n+1}}{\partial \bm{\rho}} = \Bigg[\frac{\partial \phi^r_{n+1}}{\partial \bm{\rho}}+\frac{\partial \Delta \phi^g_{n+1}}{\partial \bm{\rho}}\Bigg]:\mathbb{I} - \phi^g_{n+1}\Bigg[{\mathbb{A}^g}^{-1}\frac{\partial \mathbb{A}_g}{\partial \bm{\rho}}{\mathbb{A}^g}^{-1}\mathbb{A}_r + {\mathbb{A}^g}^{-1}\frac{\partial \mathbb{A}^r}{\partial \bm{\rho}}\Bigg]\\
+\frac{\Delta t}{\eta_i}\frac{\partial \mathbb{A}^r}{\partial \bm{\rho}}
+\frac{\partial \phi^g_{n+1}}{\partial \bm{\rho}}{\mathbb{A}_g}^{-1}\mathbb{A}_r
-\frac{\Delta t}{\eta^2_i}\mathbb{A}_r\frac{\partial \eta_i}{\partial \bm{\rho}}
\end{multline}
\begin{multline}
\frac{\partial \bm{C}_{n+1}}{\partial \bm{\rho}} = \frac{\partial \phi^g_{n+1}}{\partial \bm{\rho}}\Bigg[{\mathbb{A}_g}^{-1}\Bigg[-\mathbb{B}_r\bm{\varepsilon}^{ir}_{n}+\mathbb{B}_g\bm{\varepsilon}^{ig}_n\Bigg]\\
+\phi^g_{n+1}\Bigg[-{\mathbb{A}^g}^{-1}\frac{\partial \mathbb{A}^g}{\partial \bm{\rho}}{\mathbb{A}^g}^{-1}\{-\mathbb{B}^r\bm{\varepsilon}^{ir}_n+\mathbb{B}^g\bm{\varepsilon}^{ig}_n\} +{\mathbb{A}^g}^{-1}\{-\frac{\partial \mathbb{B}^r}{\partial \bm{\rho}}\bm{\varepsilon}^{ir}_n-\mathbb{B}^r\frac{\bm{\varepsilon}^{ir}_n}{\partial\bm{\rho}}+\frac{\partial \mathbb{B}^g}{\partial \bm{\rho}}\bm{\varepsilon}^{ig}_n+\mathbb{B}^g\frac{\partial \bm{\varepsilon}^{ig}_n}{\partial \bm{\rho}}\}\Bigg]\\
+\frac{\Delta t}{\eta_i}\Bigg[\frac{\partial \mathbb{B}^r}{\partial \bm{\rho}}\bm{\varepsilon}^{ir}_n+\mathbb{B}^r\frac{\partial \bm{\varepsilon}^{ir}_n}{\partial \bm{\rho}}\Bigg]\\
-\frac{\Delta t}{\eta^2_i}\mathbb{B}^r\bm{\varepsilon}^{ir}_n\frac{\partial \eta_i}{\partial \bm{\rho}}
-\frac{\partial\bm{ \varepsilon}^i_n}{\partial \bm{\rho}}-\frac{\partial \bm{\varepsilon}^{s}_n}{\partial \bm{\rho}}-\frac{\partial \bm{\varepsilon}^T}{\partial \bm{\rho}}
\end{multline}
The term $\frac{\partial \mathbb{B}^r}{\partial \bm{\rho}}$ is computed as:
\begin{equation}
	\frac{\partial \mathbb{B}^r}{\partial \bm{\rho}} = \{-{\mathbb{H}^r}^{-1}\frac{\partial \mathbb{H}^r}{\partial \bm{\rho}}{\mathbb{H}^r}^{-1}\}\mathbb{K}^r_{neq} +{\mathbb{H}^r}^{-1}\frac{\partial \mathbb{K}^r_{neq}}{\partial \bm{\rho}}
\end{equation}
Similar process can be adopted to compute $\frac{\partial \mathbb{B}^g}{\partial \bm{\rho}}$.

The derivatives of the strain components with respect to the design variable are given below.
\begin{multline}
	\frac{\partial \bm{\varepsilon}^{ir}_{n+1}}{\partial \bm{\rho}} = -{\mathbb{H}^r}^{-1}\frac{\partial \mathbb{H}^r}{\partial \bm{\rho}}\mathbb{H}^r\bm{\varepsilon}^{ir}_n+{\mathbb{H}^r}^{-1}\frac{\partial \bm{\varepsilon}^{ir}_n}{\partial \bm{\rho}}
	+\frac{\Delta t}{\eta^r}\{{-\mathbb{H}^r}^{-1}\frac{\partial \mathbb{H}^r}{\partial \bm{\rho}}{\mathbb{H}^r}^{-1}\mathbb{K}^r_{neq}\}\bm{\varepsilon}^r_{n+1}
	\\
	+\frac{\Delta t}{\eta^r}\{{\mathbb{H}^r}^{-1}\frac{\partial \mathbb{K}^r_{neq}}{\partial \bm{\rho}}\bm{\varepsilon}^r_{n+1} \}
	+\frac{\Delta t}{\eta^r}\{{\mathbb{H}^r}^{-1} \mathbb{K}^r_{neq}\frac{\partial \bm{\varepsilon}^r_{n+1}}{\partial \bm{\rho}} \}
	+\frac{\Delta t}{{\eta^r}^2}{\mathbb{H}^r}^{-1}\mathbb{K}^r_{neq}\bm{\varepsilon}^r_{n+1}\frac{\partial \eta^r}{\partial \bm{\rho}}
\end{multline}

\begin{multline}
	\frac{\partial \bm{\varepsilon}^i_{n+1}}{\partial \bm{\rho}} = \frac{\partial \bm{\varepsilon}^i_{n}}{\partial \bm{\rho}} +\frac{\Delta t}{{\eta^r}}\{ \mathbb{A}^r\frac{\partial \bm{\varepsilon}^r_{n+1}}{\partial \bm{\rho}} + \bm{\varepsilon}^r_{n+1}\frac{\partial \mathbb{A}^r_{n+1}}{\partial \bm{\rho}} 
	-\mathbb{B}^r\frac{\partial \bm{\varepsilon}^{ir}_{n}}{\partial \bm{\rho}} 
	-\bm{\varepsilon}^{ir}_n\frac{\partial \mathbb{B}^r}{\partial \bm{\rho}}\}\\
	-\frac{\Delta t}{{\eta^r}^2}\frac{\partial \eta^i}{\partial \bm{\rho}}\{  \mathbb{A}^r\bm{\varepsilon}^{r}_{n+1} -\mathbb{B}^r\bm{\varepsilon}^{ir}_n  \}
\end{multline}
\begin{multline}
	\frac{\partial \bm{\varepsilon}^{ig}_{n+1}}{\partial \bm{\rho}} = -{\mathbb{H}^g}^{-1}\frac{\partial \mathbb{H}^g}{\partial \bm{\rho}}\mathbb{H}^g\bm{\varepsilon}^{ig}_n+{\mathbb{H}^g}^{-1}\frac{\partial \bm{\varepsilon}^{ig}_n}{\partial \bm{\rho}}
	+\frac{\Delta t}{\eta^g}\{{-\mathbb{H}^g}^{-1}\frac{\partial \mathbb{H}^g}{\partial \bm{\rho}}{\mathbb{H}^g}^{-1}\mathbb{K}^g_{neq}\}\bm{\varepsilon}^g_{n+1}
	\\
	+\frac{\Delta t}{\eta^g}\{{\mathbb{H}^g}^{-1}\frac{\partial \mathbb{K}^g_{neq}}{\partial \bm{\rho}}\bm{\varepsilon}^g_{n+1} \}
	+\frac{\Delta t}{\eta^g}\{{\mathbb{H}^g}^{-1} \mathbb{K}^g_{neq}\frac{\partial \bm{\varepsilon}^g_{n+1}}{\partial \bm{\rho}} \}
	+\frac{\Delta t}{{\eta^g}^2}{\mathbb{H}^g}^{-1}\mathbb{K}^g_{neq}\bm{\varepsilon}^g_{n+1}\frac{\partial \eta^g}{\partial \bm{\rho}}
	\end{multline}
\begin{multline}
	\frac{\partial \bm{\varepsilon}^g}{\partial\bm{\rho}} = {\mathbb{A}^g}^{-1}\{\frac{\partial \mathbb{A}^r}{\partial \bm{\rho}}\bm{\varepsilon}^r+\mathbb{A}^r\frac{\partial \bm{\varepsilon}^r}{\partial \bm{\rho}}-\frac{\partial \mathbb{B}^r}{\partial \bm{\rho}}\bm{\varepsilon}^{ir}-\mathbb{B}^r\frac{\bm{\varepsilon}^{ir}_n}{\partial \bm{\rho}} +\frac{\partial \mathbb{B}^g}{\partial \bm{\rho}}\bm{\varepsilon}^{ig}+\mathbb{B}^g\frac{\bm{\varepsilon}^{ig}_n}{\partial \bm{\rho}}\}
	\\+ \{-{\mathbb{A}^g}^{-1}\frac{\partial \mathbb{A}^g}{\partial \bm{\rho}}{\mathbb{A}^g}^{-1}\}\mathbb{A}^r\bm{\varepsilon}^r_{n+1}
	-\mathbb{B}^r\bm{\varepsilon}^{ir}_n
	+\mathbb{B}^g\bm{\varepsilon}^{ig}_n
\end{multline}
The derivative of the stored strain with respect to the design variable can be written as:
\begin{equation}
	\frac{\partial \bm{\varepsilon}^s_{n+1}}{\partial \bm{\rho}} = 	\frac{\partial \bm{\varepsilon}^s_n}{\partial \bm{\rho}} + \Delta \phi^g\frac{\partial \bm{\varepsilon}^r_{n+1}}{\partial \bm{\rho}} + \frac{\partial \Delta \phi^g_{n+1}}{\partial \bm{\rho}}\bm{\varepsilon}^r_{n+1}
\end{equation}

To evaluate the \emph{adjoint} vectors, it is required to capture the explicit dependence of the residual for the $k^{th}$ time step on the displacement of the $i^{th}$ time step i.e $\frac{\partial \bm{R}_{k}}{\partial \bm{u}_{i}}$. These terms are referred to as the ``coupling" terms. Finding the $\frac{\partial \bm{R}_{k}}{\partial \bm{u}_{i}}$ terms are more involved since at each time step there is an exponential growth of terms from the previous time step. For example, let us evaluate the term $\frac{\partial \bm{R}_{n+1}}{\partial \bm{u}_{n-1}}$. The coupling term $\frac{\partial \bm{R}_{n+1}}{\partial \bm{u}_{n-1}}$ is proportional to $\frac{\partial \bm{R}_{n+1}}{\partial \bm{\varepsilon}_{n-1}}$, since strain is a linear function of displacement ($u$). We can use the chain rule to write 
\begin{align}
	\begin{split}
		\frac{\partial \bm{R}_{n+1}}{\partial \bm{u}_{n-1}}&\propto\frac{\partial \bm{R}_{n+1}}{\partial \bm{\varepsilon}_{n-1}}\approx\underbrace{\frac{\partial \bm{R}_{n+1}}{\partial \bm{\varepsilon}^{(r)}_{n-1}}}_\text{term I}\overbrace{\frac{\partial \bm{\varepsilon}^{(r)}_{n-1}}{\partial \bm{\varepsilon}_{n-1}}}^{\text{term II}}
	\end{split}
	\label{proportional}
\end{align}
Equation {(\ref{proportional})} gets contributions from \emph{term I} and \emph{term II}. The parameter $\bm{R}_{n+1}$ which represents the residual, obtained during the forward analysis, is given by Equation {(\ref{residual eqn})} which has seven terms. The \emph{term I} can be further written as:

\begin{equation}
	\frac{\partial \bm{R}_{n+1}}{\partial \bm{\varepsilon}^{(r)}_{n-1}} = \frac{\partial \bm{F}^{int}_{n+1}}{\partial \bm{\varepsilon}^{(r)}_{n-1}} = \mathbb{A}^r{\mathbb{D}_{n+1}}^{-1}\frac{\partial \bm{C}_{n+1}}{\partial \bm{\varepsilon}^{(r)}_{n-1}} - \mathbb{B}^r\frac{\partial \bm{\varepsilon}^{ir}_n}{\partial \bm{\varepsilon}^{(r)}_{n-1}}
\end{equation}
The term $\frac{\partial \bm{C}_{n+1}}{\partial \bm{\varepsilon}^{(r)}_{n-1}} $ can be written as :
\begin{equation}
\frac{\partial \bm{C}_{n+1}}{\partial \bm{\varepsilon}^{(r)}_{n-1}} = C^{ir}_0\frac{\partial \bm{\varepsilon}^{ir}_{n}}{\partial \bm{\varepsilon}^{(r)}_{n-1}} +C^{ig}_0\frac{\partial \bm{\varepsilon}^{ig}_{n}}{\partial \bm{\varepsilon}^{(r)}_{n-1}} +C^{i}_0\frac{\partial \bm{\varepsilon}^{i}_{n}}{\partial \bm{\varepsilon}^{(r)}_{n-1}}+C^{is}_0\frac{\partial \bm{\varepsilon}^{is}_{n}}{\partial \bm{\varepsilon}^{(r)}_{n-1}}
\end{equation}
Here, $C^{\square}_0$ are appropriate constants for the specific strain derivative terms.

\begin{figure}[H]
	\centering
	\includegraphics[width= 14cm,trim={0cm 0cm 0cm 0cm },clip]{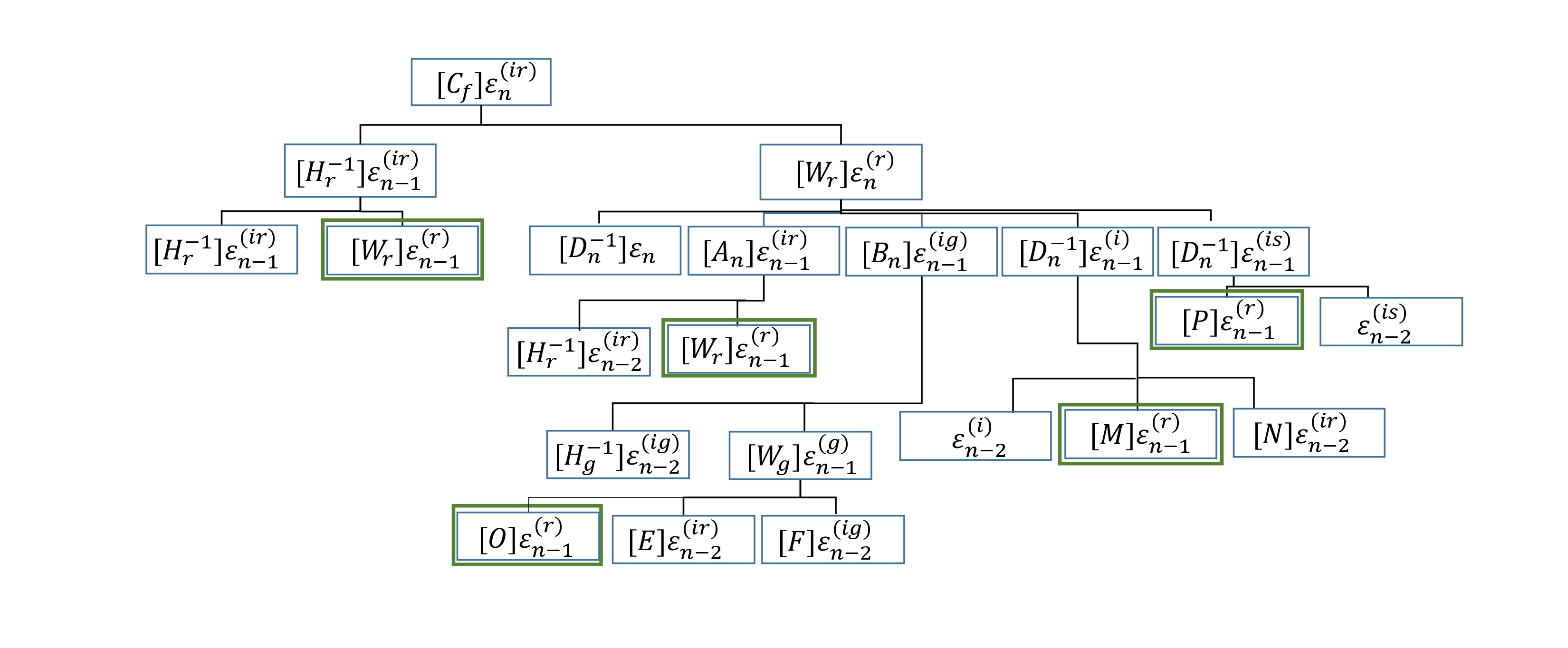}
	\caption{Tracking $\frac{\partial \bm{\varepsilon}^{(ir)}_{n}}{\partial \bm{\varepsilon}^{(r)}_{n-1}}$ terms in time}
	\label{chase}
\end{figure}

 Each of the terms, at a particular time step, is dependent not only on the current time step of the evaluation but also on the previous time step as shown in Equation {(\ref{Contributions}). For example, if we calculate the coupling coefficients from the second term, $\int_{\Omega}^{} \bm{B}^T\mathbb{X}^{(r)}_{n+1}\bm{\varepsilon}^{(ir)}_{n} dv$, of the residual equation, and track the evolution of the term in time, we will get the map as shown in Figure {(\ref{chase})}. The coefficient $C^{ir}_0 = C_f$ is defined as}
\begin{align*}
	\begin{split}
		C_f &= \bm{B}^T\mathbb{X}^{(r)}_{n+1}\\
	\end{split}
	\label{cf}
\end{align*}
The terms $\mathbb{A}_n$ and $\mathbb{B}_n$ are given by
\begin{equation}
	\begin{split}
		\mathbb{A}_n &= \mathbb{D}^{-1}_n\Bigg[-\phi^g_n\mathbb{A}^{-1}_g\mathbb{B}_r+\frac{\Delta t}{\eta_i}\mathbb{B}_r\Bigg] \\
		\mathbb{B}_n & =\mathbb{D}^{-1}_n\Bigg[\phi^g_n\mathbb{A}^{-1}_g\mathbb{B}_g\Bigg] 
	\end{split}
	\label{A_n B_n}
\end{equation}
If we collect the terms to evaluate $\frac{\partial \bm{\varepsilon}^{(ir)}_{n}}{\partial \bm{\varepsilon}^{(r)}_{n-1}}$, we get
\begin{align}
	\begin{split}
		\frac{\partial \bm{\varepsilon}^{(ir)}_{n}}{\partial \bm{\varepsilon}^{(r)}_{n-1}}&=\Bigg[\mathbb{H}^{-1}_r\mathbb{W}_r+\mathbb{W}_r\mathbb{A}_n\mathbb{W}_r+\mathbb{W}_r\mathbb{B}_n\mathbb{W}_g\mathbb{O}+\mathbb{W}_r\mathbb{D}^{-1}_n\mathbb{M}+\mathbb{W}_r\mathbb{D}^{-1}_n\mathbb{P}\Bigg]
	\end{split}
	\label{deirndern-1}
\end{align}
Equation {(\ref{deirndern-1})} represents \emph{term I} in terms of $\varepsilon^{(ir)}_n$. A similar procedure is adopted for all the other six terms present in the Equation {(\ref{residual eqn})} to make a total of twenty-three terms for the coupling term $\frac{\partial \bm{R}_{n+1}}{\partial \bm{u}_{n-1}}$.
\begin{figure}[H]
	\centering
	\includegraphics[width= 14cm,trim={0cm 0cm 0cm 0cm },clip]{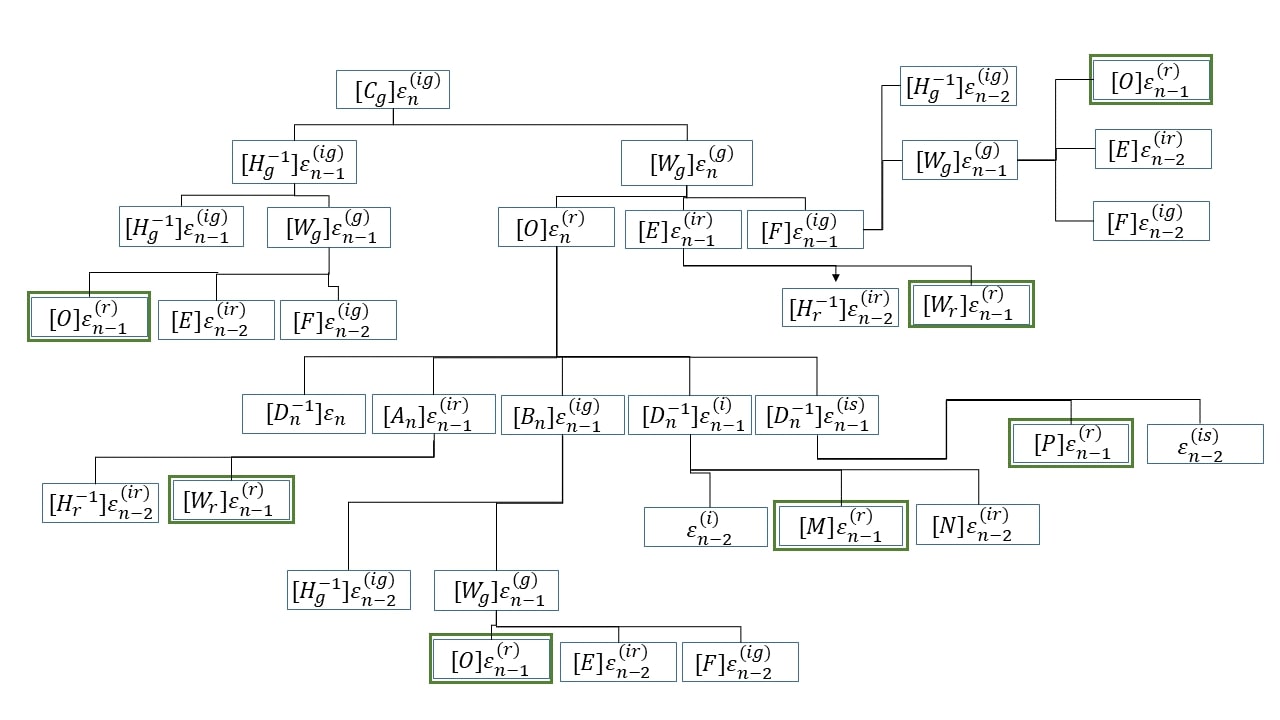}
	\caption{Tracking $\frac{\partial \bm{\varepsilon}^{(ig)}_{n}}{\partial \bm{\varepsilon}^{(r)}_{n-1}}$ terms in time}
	\label{e ig evolution}
\end{figure}
\begin{multline}
		\frac{\partial \bm{\varepsilon}^{(ig)}_{n}}{\partial \bm{\varepsilon}^{(r)}_{n-1}} =\Bigg[\mathbb{H}^{-1}_g\mathbb{W}_g+\mathbb{O}_r+\mathbb{W}_g\mathbb{O}\mathbb{A}_n\mathbb{W}_r+\mathbb{W}_g\mathbb{O}\mathbb{B}_n\mathbb{W}_g\mathbb{O}+\mathbb{W}_g\mathbb{O}\mathbb{D}^{-1}_n\mathbb{M}+\mathbb{W}_g\mathbb{O}\mathbb{D}^{-1}_n\mathbb{P} \\+\mathbb{W}_g\mathbb{E}\mathbb{W}_r+\mathbb{W}_g\mathbb{F}\mathbb{W}_g\mathbb{O}\Bigg]
	\label{deigndern-1}
\end{multline}
\begin{figure}[H]
	\centering
	\includegraphics[width= 14cm,trim={0cm 0cm 0cm 0cm },clip]{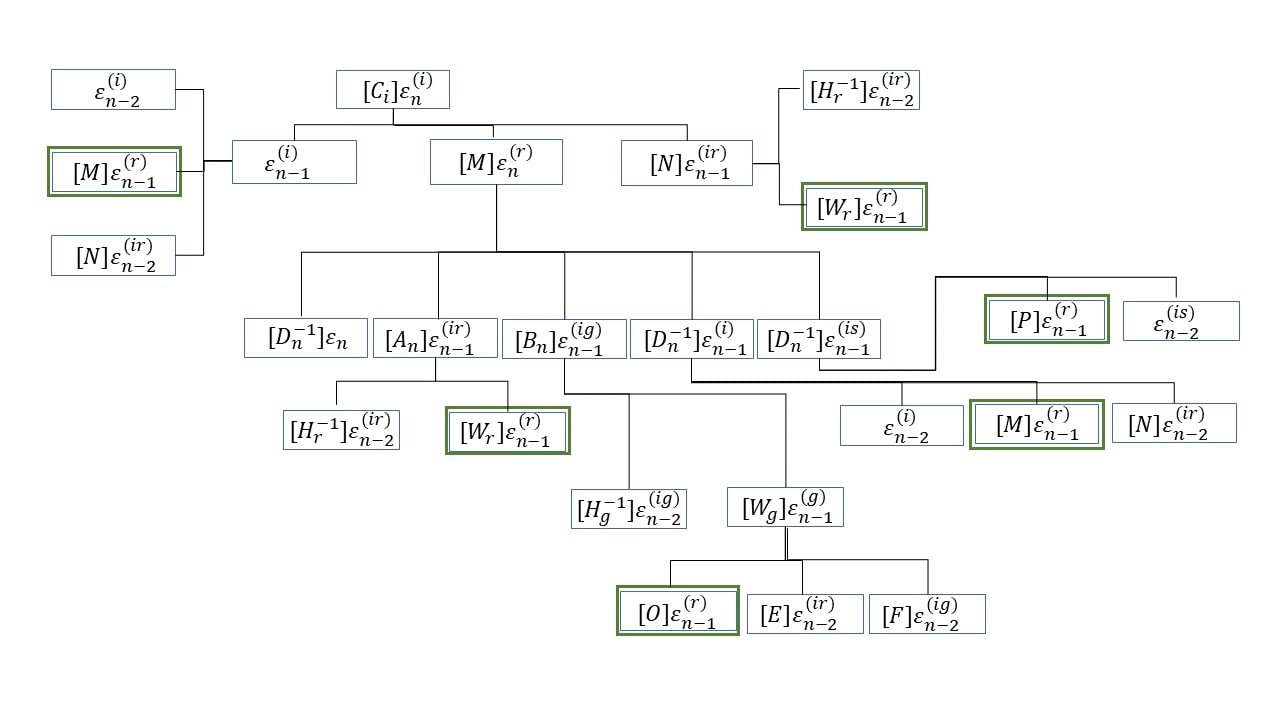}
	\caption{Tracking $\frac{\partial \bm{\varepsilon}^{(i)}_{n}}{\partial \bm{\varepsilon}^{(r)}_{n-1}}$ terms in time}
	\label{e i evolution}
\end{figure}
\begin{multline}
	\frac{\partial \bm{\varepsilon}^{(i)}_{n}}{\partial \bm{\varepsilon}^{(r)}_{n-1}} =\Bigg[\mathbb{M}+\mathbb{M}\mathbb{A}_n\mathbb{W}_r+\mathbb{M}\mathbb{B}\mathbb{W}_g\mathbb{O}+\mathbb{M}\mathbb{D}^{-1}_n\mathbb{M}+\mathbb{M}\mathbb{D}^{-1}_n\mathbb{P}+\mathbb{N}\mathbb{W}_r\Bigg]
	\label{deindern-1}
\end{multline}
\begin{figure}[H]
	\centering
	\includegraphics[width= 14cm,trim={0cm 5cm 0cm 0cm },clip]{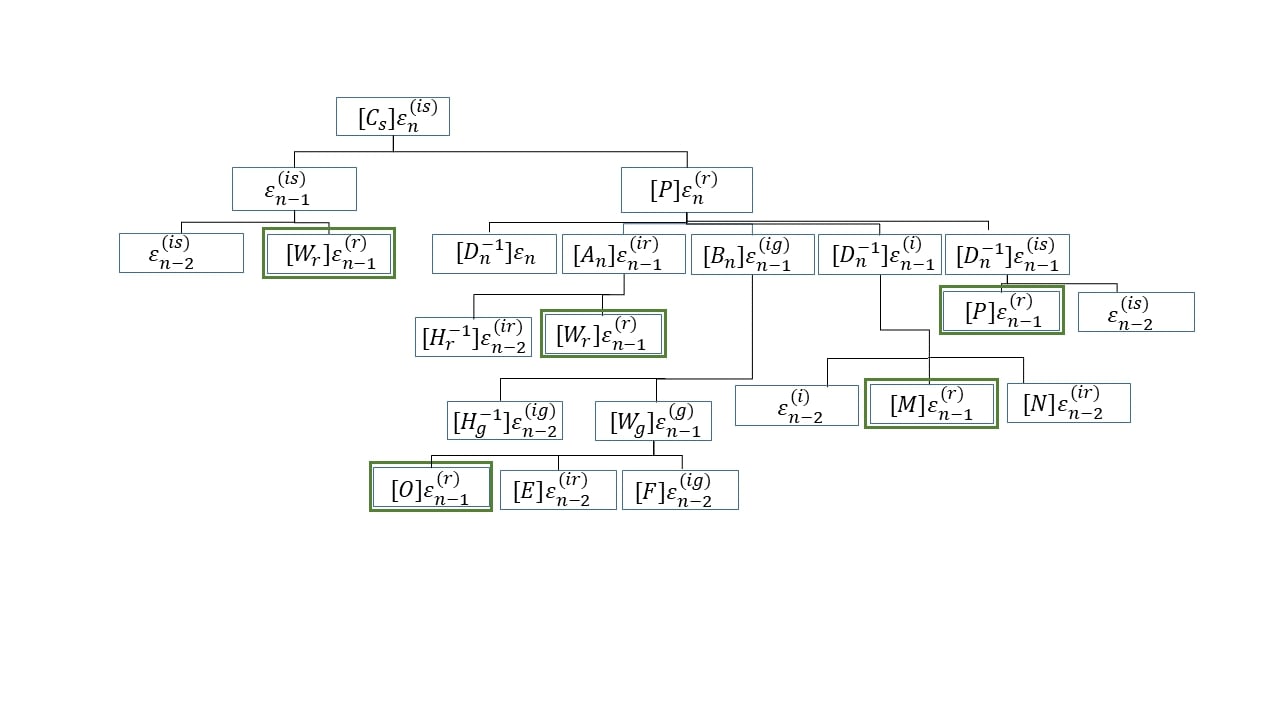}
	\caption{Tracking $\frac{\partial \bm{\varepsilon}^{(is)}_{n}}{\partial \bm{\varepsilon}^{(r)}_{n-1}}$ terms in time}
	\label{e is evolution}
\end{figure}
\begin{multline}
	\frac{\partial \bm{\varepsilon}^{(is)}_{n}}{\partial \bm{\varepsilon}^{(r)}_{n-1}} =\Bigg[\mathbb{W}_r+\mathbb{P}\mathbb{A}_n\mathbb{W}_r+\mathbb{P}\mathbb{B}_n\mathbb{W}_g\mathbb{O}+\mathbb{P}\mathbb{D}^{-1}_n\mathbb{M}+\mathbb{P}\mathbb{D}^{-1}_n\mathbb{P}\Bigg]
	\label{deisdern-1}
\end{multline}

The computation of \emph{term II} is straightforward and is given by
\begin{equation}
	\frac{\partial \bm{\varepsilon}^{(r)}_{n-1}}{\partial \bm{\varepsilon}_{n-1}} = {\mathbb{D}^{-1}_{n-1}}
\end{equation}
Capturing the evolution of all the components required to accurately calculate the sensitivities makes the this process computationally expensive and a highly time consuming procedure. The time taken increases exponentially with the total number of time steps required to simulate the thermo-mechanical cycle of the SMP increases. The function and the recursive algorithm used to compute the \{$\frac{\partial R_k}{\partial u_i}$\} terms for the total sensitivity analysis are shown in {\href{algo-2}{Algorithm 2}} and {\href{algo-3}{Algorithm 3}}. Note that for the recursive algorithm shown in {\href{algo-3}{Algorithm 3}}, parameters $k$ and $i$ represent the time-steps. Here, the functions \texttt{func\_eir, func\_eig, func\_is, func\_i} are programmable versions of $\bm{\varepsilon^{(ir)}}, \bm{\varepsilon^{(ig)}}, \bm{\varepsilon^{(is)}}, \bm{\varepsilon^{(i)}}$, shown in Equation 9\cite{bhattacharyya2021topology}, implemented for the $k^{th}$ step. The variable $[M]$ is a collection of parameters representing the intrinsic material properties. The function $f$ represents a general function manipulating its inputs and giving a desired output.

\begin{algorithm}[H]
	\DontPrintSemicolon
	\SetKwFunction{FRecurs}{FnRecursive}
	\SetKwFunction{Feir}{func\_eir}%
	\SetKwFunction{Fer}{func\_er}%
	\SetKwFunction{Feig}{func\_eig}%
	\SetKwFunction{Feg}{func\_eg}%
	\SetKwFunction{Fis}{func\_is}%
	\SetKwFunction{Fi}{func\_i}%
	\SetKwFunction{dfdu}{sens\_partI}%
	\SetKwFunction{func}{func\_dfdu}%
	\dfdu{$k,i,M$}\;
	$C_f = f(M)$\tcc*{compute external variable $C_f$ as a function of $M$}
	\tcc{call individual recursive functions}
	$\frac{\partial \varepsilon^{(ir)}_k}{\partial \varepsilon^{(r)}_i}\leftarrow$ \Feir{$C_f,k,i,M$}\\
	$\frac{\partial \varepsilon^{(ig)}_k}{\partial \varepsilon^{(r)}_i}\leftarrow$ \Feig{$C_f,k,i,M$}\\
	$\frac{\partial \varepsilon^{(is)}_k}{\partial \varepsilon^{(r)}_i}\leftarrow$ \Fis{$C_f,k,i,M$}\\
	$\frac{\partial \varepsilon^{(i)}_k}{\partial \varepsilon^{(r)}_i}\leftarrow$ \Fi{$C_f,k,i,M$}\;
	term I = $f\Bigg(\frac{\partial \varepsilon^{(ir)}_k}{\partial \varepsilon^{(r)}_i},\frac{\partial \varepsilon^{(ig)}_k}{\partial \varepsilon^{(r)}_i},\frac{\partial \varepsilon^{(is)}_k}{\partial \varepsilon^{(r)}_i},\frac{\partial \varepsilon^{(i)}_k}{\partial \varepsilon^{(r)}_i}\Bigg)$\tcc*{term I of (\ref{proportional}) is calculated using the output of the individual recursive functions}
	term II = $f\Big(\frac{\partial \varepsilon^{(r)}_i}{\partial \varepsilon_i}\Big)$\tcc*{term II of (\ref{proportional}) is calculated }
	\textbf{Return:}\vspace{2mm}{$\frac{\partial R_k}{\partial u_i} \leftarrow f(\text{term I, term II})$}
	\caption{psuedocode to calculate the terms $\frac{\partial R_k}{\partial u_i}$ for the sensitivity evaluation \label{algo-2} }
\end{algorithm}
The individual functions have similar structures and one such function \texttt{func\_eir} has been shown in details in {\href{algo-3}{Algorithm 3}}.

\begin{figure}[H]
	\centering
	\includegraphics[width= 11cm,trim={0cm 0cm 0cm 0cm },clip]{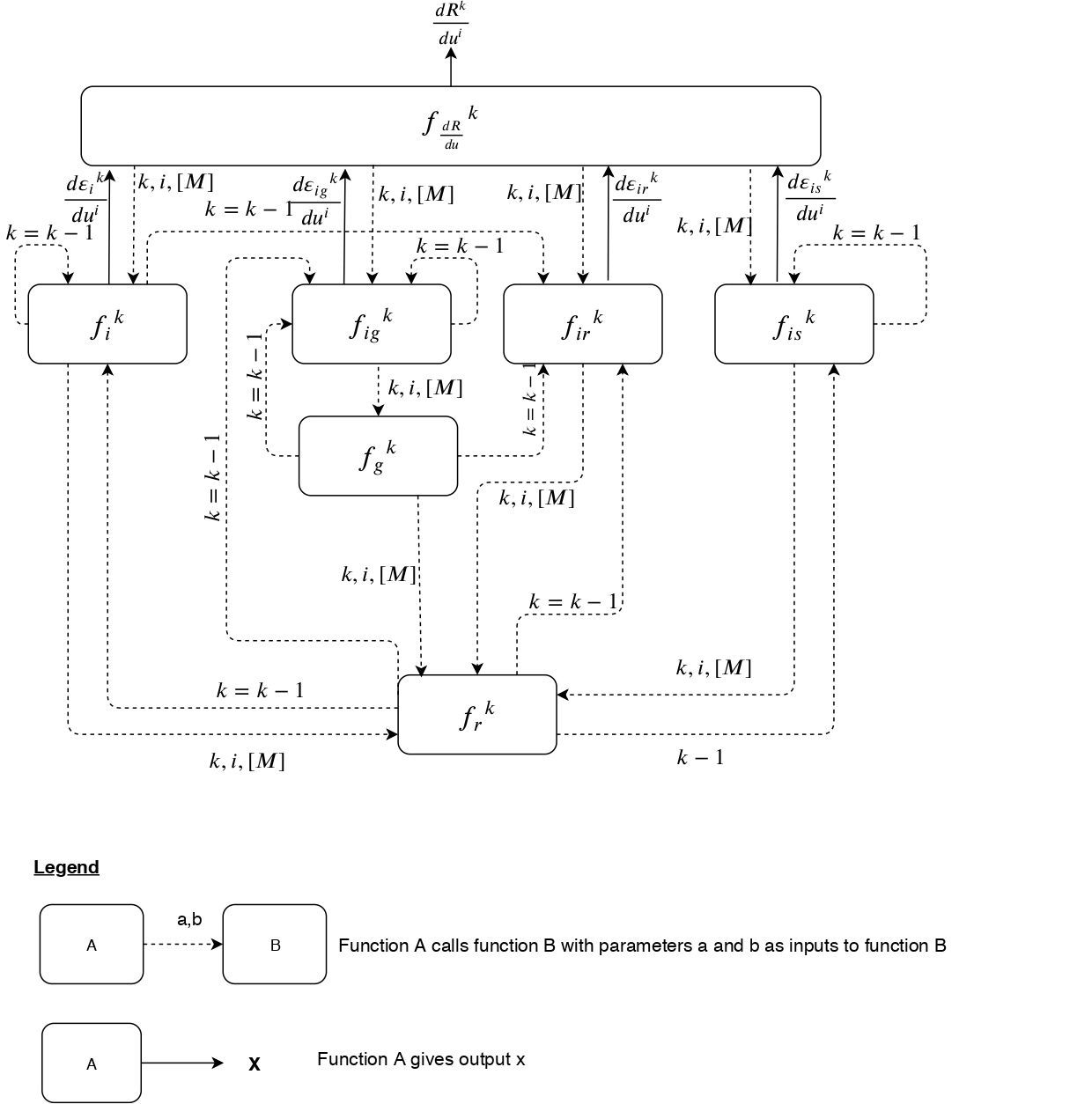}
	\caption{Recursive algorithm to calculate the terms $\frac{dR^k}{du^i}$ for the sensitivity evaluation}
	\label{algo}
\end{figure}

\begin{algorithm}[H]
	\DontPrintSemicolon
	\SetKwFunction{FRecurs}{FnRecursive}
	\SetKwFunction{Feir}{func\_eir}%
	\SetKwFunction{Fer}{func\_er}%
	\SetKwFunction{Feig}{func\_eig}%
	\SetKwFunction{Feg}{func\_eg}%
	\SetKwFunction{Fis}{func\_is}%
	\SetKwFunction{Fi}{func\_i}%
	\SetKwFunction{dfdu}{sens\_partI}%
	\SetKwFunction{func}{func\_dfdu}%
	\Feir{$C_f, k,i,M$}\;
	$CI = \mathbb{H}^{-1}_r$\tcc*{compute internal variable $CI$}
	$\frac{\partial \varepsilon^{(r)}_k}{\partial \varepsilon^{(r)}_i}\leftarrow$\Fer{$CI,k, i, M$}\tcc*{call function which tracks evolution of strain variables as shown in Equation (\ref{deirndern-1})}
	\If{$k>i$}{$\frac{\partial \varepsilon^{(ir)}_{k-1}}{\partial \varepsilon^{(r)}_i}\leftarrow$\Feir{$C_f\times CI, k-1,i, M$}}\tcc*{call itself with $k=k-1$ }

	\textbf{Return:}\vspace{2mm}{$\frac{\partial \varepsilon^{(r)}_k}{\partial \varepsilon^{(r)}_i}+\frac{\partial \varepsilon^{(ir)}_{k-1}}{\partial \varepsilon^{(r)}_i}$\tcc*{output}}
	\caption{Recursive algorithm to capture strain evolution with time for the sensitivity evaluation \label{algo-3} }
\end{algorithm}
Figure \ref{algo} gives an overview of the recursive algorithm implemented to compute $\frac{\partial \bm{R}_k}{\partial \bm{u}_i}$ for the sensitivity analysis formulation. 

\begin{figure}[H]
	\centering
	\includegraphics[width= 8cm,trim={0cm 0cm 0cm 0cm },clip]{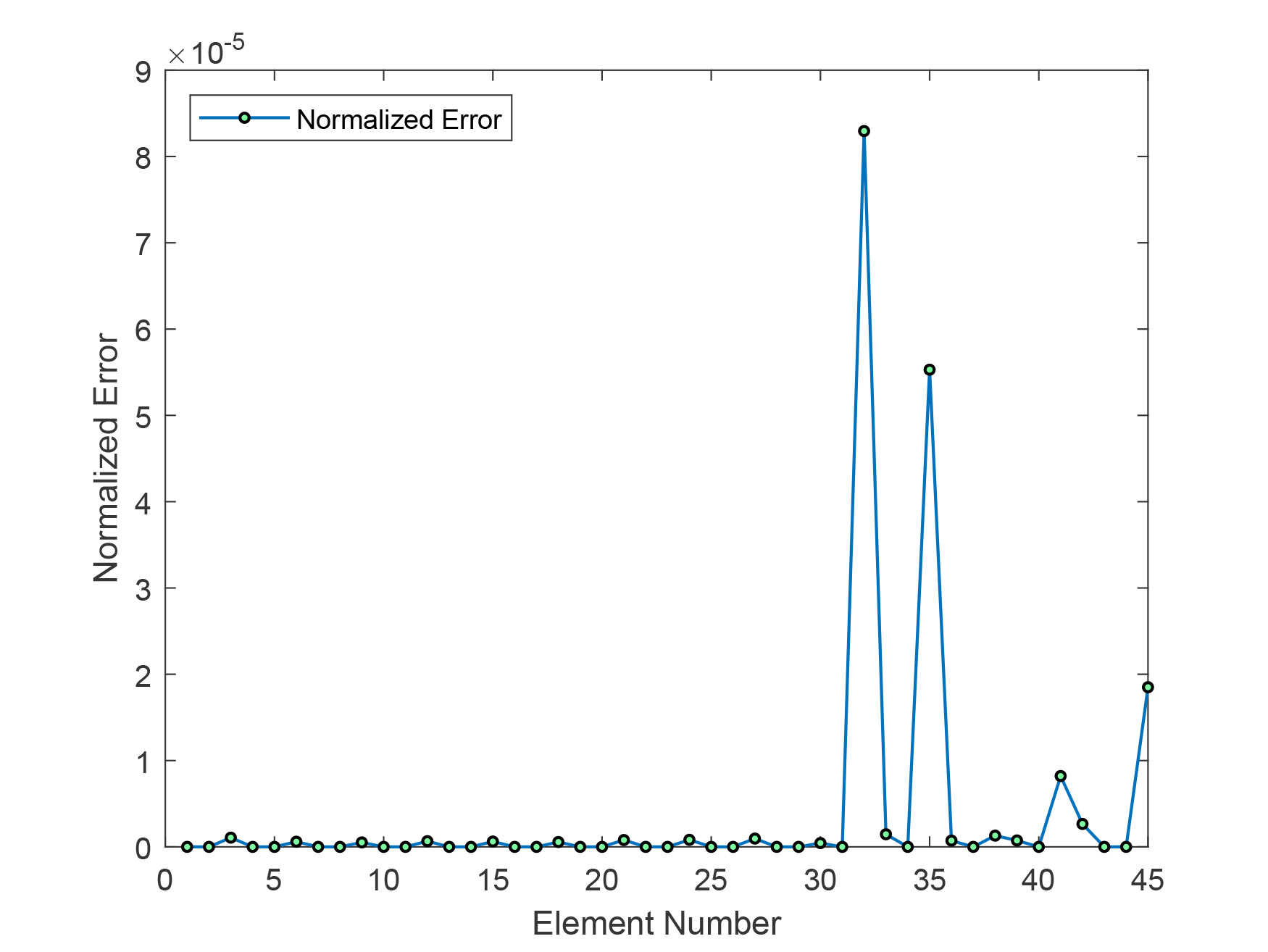}
	\caption{Comparison between the sensitivity values evaluated through the finite-difference scheme and the adjoint formulation}
	\label{adj-fd}
\end{figure}

\begin{figure}[H]
	\centering
	\includegraphics[width= 8cm,trim={0cm 0cm 0cm 0cm },clip]{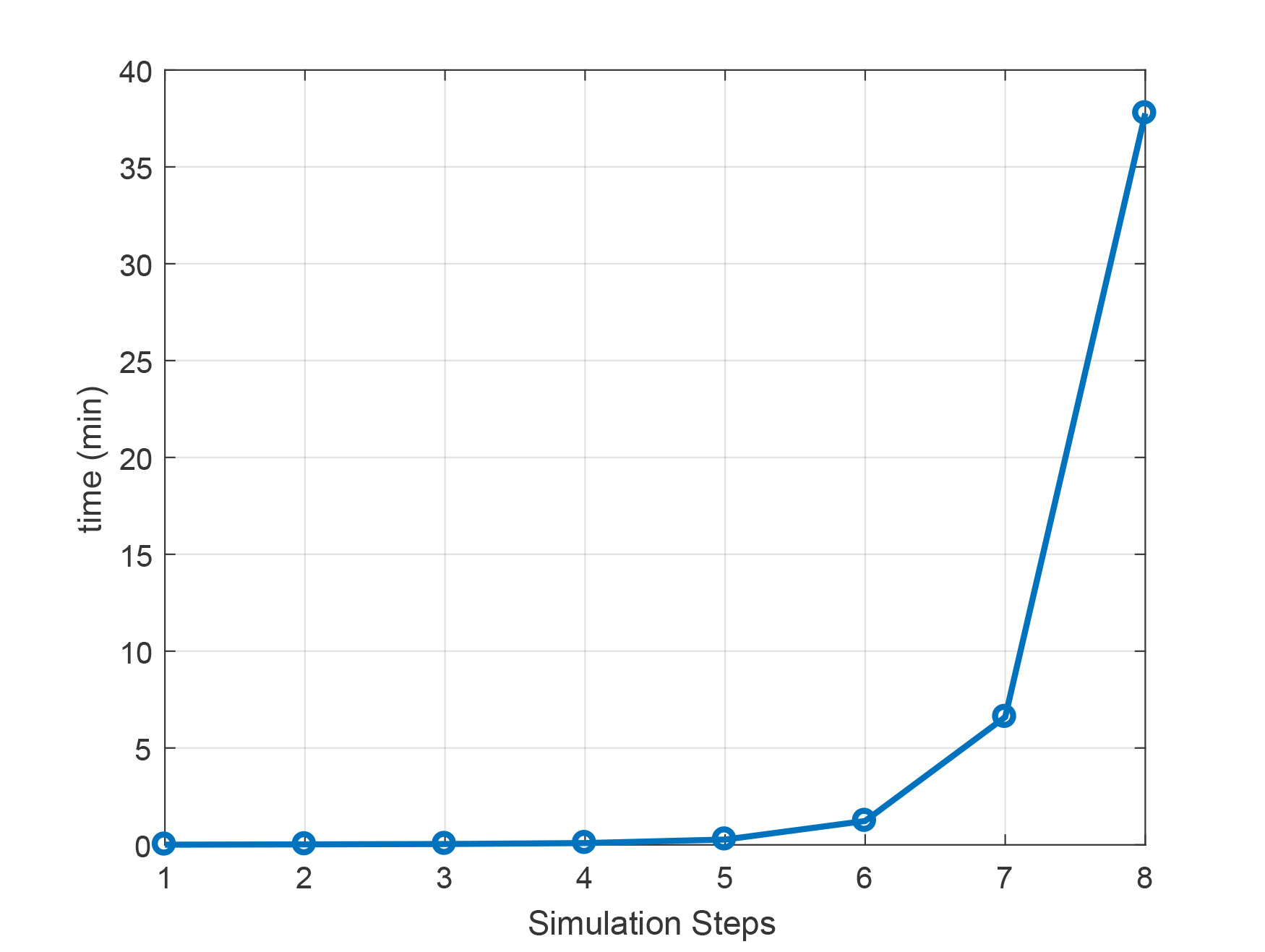}
	\caption{Computation time required for tracking $\frac{\partial \bm{\varepsilon}^{ir}_{n}}{\partial \bm{\varepsilon}^r_{n-1}}$ terms}
	\label{steps_vs_time}
\end{figure}

To verify the implementation of the sensitivity analysis, the design domain shown in Figure {(\ref{sens domain})} is discretized with a coarse mesh of 45 elements. The structure is initialised with an uniform distribution of design variable $\bm{\rho}=0.3$. It was then subjected to an axial stretching load $\bm{F}=0.025\; N$ during the cooling phase of the thermo-mechanical cycle. The load was removed during the relaxation and heating phase of the thermo-mechanical programming cycle. The function of interest is the tip displacement $\bm{u}^M_a$ as shown in Equation {(\ref{objective function})}. In this case, the parameter $a$ is the $y-$ degree-of-freedom of the node shown in Figure {(\ref{sens domain})} and $M$ is the time step at the end of the Step-III of the thermo-mechanical programming cycle. The material parameters used for this analysis is same as listed in Table 1\cite{bhattacharyya2021topology}. The adjoint method and the forward difference method were used to evaluate the derivative of the tip displacement with respect to the mixing ratio of each element.  Figure {(\ref{adj-fd})} shows the normalised error of the sensitivity values obtained by the finite-difference approach and the adjoint sensitivity analysis. The normalised error (NE) for each element is evaluated as
\begin{equation}
	NE = \Bigg|\frac{adjoint - FD}{FD}\Bigg|
\end{equation}
Note that for elements where the sensitivity is at or near zero, we have omitted the normalized error to avoid the indication of an artificially high error due to an extremely small denominator.  The displacement obtained at the end of Step-III was $-0.0130\;mm$. The maximum error between these values was found to be 2.6$\times$ $10^{-7}$. This established that the framework developed can successfully compute the sensitivities for SMP materials with a high degree of accuracy.
Figure {(\ref{steps_vs_time})} shows the time required to calculate $\frac{\partial \bm{R}_{n+1}}{\partial \bm{u}_{n-7}}$, the contribution of a total of 8 simulation steps , for a finite-element mesh of 50 elements by a single processor. As we can see, just using eight steps to simulate the entire SMP thermo-mechanical programming cycle, even for a coarse mesh can incur high computational costs.  
This result motivated the development of PETSc-based parallel implementation of the FEA and sensitivity evaluation framework using CPUs on the \emph{Golub Cluster} at the University of Illinois. Since the bottleneck for the entire algorithm is the sensitivity evaluation and particularly the time-dependent algorithm, the parallelization is done with the objective of distributing the elements onto the processors such that each processor has the optimum number of elements for efficient computations. A total of 144 processors (6 nodes with 24 processors each) were utilized for generating the 2D results. For the 3D optimization implementation, a total of 250 processors (10 nodes with 25 processors each) were utilized.  

\begin{table}[H]
	\small
	\caption{Sensitivity values evaluated through the adjoint formulation and the finite difference method.}
	\label{sens table 1}
	\begin{tabular}{|l|l|l|l|}
		\hline
		
		Element No. & Adjoint Sensitivities & Finite Difference Sensitivities & Normalized Error($\times10^{-6}$)                          \\ \hline
		\hline
		1           & -0.2416477            & -0.2416476                      & 0.482 \\ \hline
		2           & -0.0000000            & -0.0000001                      & --                                     \\ \hline
		3           & 0.2416477             & 0.2416475                       & 1.07 \\ \hline
		4           & -0.2543351            & -0.2543351                      & 0.00                                     \\ \hline
		5           & 0.0000000             & 0.0000000                       & 0.00                                     \\ \hline
		6           & 0.2543351             & 0.2543350                       & 0.599 \\ \hline
		7           & -0.2375154            & -0.2375154                      & 0.00                                     \\ \hline
		8           & 0.0000000             & 0.0000000                       & 0.00                                     \\ \hline
		9           & 0.2375154             & 0.2375153                       & 0.516  \\ \hline
		10          & -0.2225283            & -0.2225282                      & 0.376  \\ \hline
		11          & 0.0000000             & 0.0000000                       & 0.00                                     \\ \hline
		12          & 0.2225283             & 0.2225281                       & 0.661  \\ \hline
		13          & -0.2038087            & -0.2038086                      & 0.359 \\ \hline
		14          & 0.0000000             & 0.0000000                       & 0.00                                     \\ \hline
		15          & 0.2038086             & 0.2038085                       & 0.619 \\ \hline
		16          & -0.1844918            & -0.1844917                      & 0.539 \\ \hline
		17          & -0.0000001            & -0.0000001                      & 0.00                                     \\ \hline
		18          & 0.1844916             & 0.1844915                       & 0.563 \\ \hline
		19          & -0.1650572            & -0.1650571                      & 0.244 \\ \hline
		20          & -0.0000010            & -0.0000010                      & 0.00                                     \\ \hline
		21          & 0.1650567             & 0.1650565                       & 0.793 \\ \hline
		22          & -0.1456293            & -0.1456293                      & 0.00                                     \\ \hline
		23          & -0.0000042            & -0.0000041                      & --                                    \\ \hline
		24          & 0.1456292             & 0.1456291                       & 0.818 \\ \hline
		25          & -0.1262059            & -0.1262059                      & 0.00                                     \\ \hline
		26          & -0.0000104            & -0.0000104                      & 0.00                                     \\ \hline
		27          & 0.1262118             & 0.1262119                       & 0.958\\ \hline
		28          & -0.1067716            & -0.1067717                      & 0.454  \\ \hline
		29          & -0.0000017            & -0.0000017                      & 0.00                                     \\ \hline
		30          & 0.1068102             & 0.1068101                       & 0.441 \\ \hline
		31          & -0.0873091            & -0.0873091                      & 0.00                                     \\ \hline
		32          & 0.0001378             & 0.0001379                       & 82.9                                     \\ \hline
		33          & 0.0874321             & 0.0874320                       & 1.46 \\ \hline
		34          & -0.0678103            & -0.0678104                      & 0.276  \\ \hline
		35          & 0.0008085             & 0.0008084                       & 55.2                                     \\ \hline
		36          & 0.0680543             & 0.0680542                       & 0.724\\ \hline
		37          & -0.0488513            & -0.0488514                      & 1.12                                     \\ \hline
		38          & 0.0027865             & 0.0027864                       & 1.30                                     \\ \hline
		39          & 0.0480513             & 0.0480513                       & 0.00                                     \\ \hline
		40          & -0.0302951            & -0.0302951                      & 0.00                                     \\ \hline
		41          & 0.0042528             & 0.0042528                       & 0.00                                     \\ \hline
		42          & 0.0250887             & 0.0250886                       & 2.65                                     \\ \hline
		43          & -0.0388250            & -0.0388251                      & 0.415  \\ \hline
		44          & -0.0123738            & -0.0123738                      & 0.00                                     \\ \hline
		45          & 0.0027180             & 0.0027181                       & 18.5   \\ \hline
	\end{tabular}
\end{table}

\bibliographystyle{plain}


\end{document}